\documentclass[structabstract]{aa}
\usepackage{graphicx}
\usepackage{txfonts}
\begin{document}
   \title{The correlation of fractal structures in the photospheric and the coronal magnetic field}


   \author{M. Dimitropoulou,
          \inst{1}
          M. Georgoulis,
          \inst{2}
          H. Isliker,
          \inst{3}
          L. Vlahos,
          \inst{3}
          A. Anastasiadis,
          \inst{4}
          D. Strintzi
          \inst{5}
          \and
          X. Moussas\inst{1}
          }
   \institute{University of Athens, Department of Physics, GR-15483, Athens, Greece\\
              \email{michaila.dimitropoulou@nsn.com}
   \and
              Johns Hopkins University Applied Physics Laboratory, 11100 Johns Hopkins Road, Laurel, MD-20723-6099, USA\\
   \and
              University of Thessaloniki,Department of Physics, GR-54006, Thessaloniki, Greece\\
   \and
              Institute for Space Applications and Remote Sensing, National Observatory of Athens, GR-15236 Penteli, Greece\\
   \and
              National Technical University of Athens, GR-15773, Athens, Greece
             }




  \abstract
   {This work examines the relation between the fractal
    properties of the photospheric magnetic patterns
    and those of the coronal magnetic fields in solar active regions.}
   {We investigate whether there is any correlation between the fractal dimensions
   of the photospheric structures and the magnetic discontinuities formed in the corona.}
   {To investigate the connection
    between the photospheric and coronal complexity,
    we used a nonlinear force-free extrapolation method that reconstructs the $3d$ magnetic fields
    using $2d$ observed vector magnetograms as boundary conditions.
    We then located the magnetic discontinuities, which are considered as
    spatial proxies of reconnection-related instabilities.
    These discontinuities form well-defined volumes,
    called here unstable volumes. We calculated the fractal dimensions of these unstable volumes and
    compared them to the fractal dimensions of the boundary vector magnetograms.}
   {Our results show no correlation between the fractal dimensions
    of the observed $2d$ photospheric structures and the extrapolated unstable volumes
    in the corona, when nonlinear force-free extrapolation is used. This result is independent
    of efforts to (1) bring the photospheric magnetic fields closer to a nonlinear force-free equilibrium
    and (2) omit the lower part of the modeled magnetic field volume that is almost completely filled by
    unstable volumes. A significant correlation between the fractal dimensions of the
    photospheric and coronal magnetic features is only observed at the zero level (lower limit) of approximation of a
    current-free (potential) magnetic field extrapolation.}
   {We conclude that the complicated transition from photospheric non-force-free fields to
   coronal force-free ones hampers any direct correlation between the fractal dimensions of the $2d$ photospheric
   patterns and their $3d$ counterparts in the corona at the nonlinear force-free limit, which can be considered
   as a second level of approximation in this study. Correspondingly, in the zero and first levels of approximation, namely, the potential and linear force-free extrapolation, respectively, we reveal a significant correlation between the fractal dimensions of the photospheric and coronal structures, which can be attributed to the lack of electric currents or to their purely field-aligned orientation.}

   \keywords{Sun: corona --
                Sun: flares --
                Sun: photosphere
               }
\titlerunning{Fractal correlation between the photosphere and the corona}
\authorrunning{Dimitropoulou et al.}
   \maketitle
%

\section{Introduction}

Solar active regions (ARs) have attracted the interest of many researchers
over the years, since they are connected with solar energetic events, such as solar
flares and CMEs. The relation of AR magnetic complexity  with their flare productivity
has been lately widely investigated (\cite{meu04,mca05,geo08}).
There are many measures of the ARs' complexity,
such as their size distribution (\cite{har93}, \cite{meu99}), the length of the main polarity inversion line
(\cite{fal06}) or the magnetic flux along the polarity inversion line (\cite{sch07}). A common approach
nevertheless relies on fractal analysis because the
fractal dimension indicates the self-similarity of a structure over several size scales.
The sizes of AR magnetic fields  are known to
display power-law distributions
(\cite{har93}, \cite{abr05}), meaning that the very nature of AR magnetic fields that indicates fractal analysis as a suitable tool
for AR complexity determination.

A variety of fractal analysis methods are available nowadays. A basic classification of these
methods distinguishes between monofractal and multifractal techniques. An overview of monofractal algorithms
has been provided by McAteer et al. (2005), who indicates the perimeter area,
the linear size area, and the box counting methods as the most important monofractal methodologies. The pros and cons
of these techniques, as well as the resulting discrepancies are discussed in the same work.
These techniques have been widely used in the literature:
perimeter area technique (\cite{rou87}; \cite{hir97};
\cite{meu99}; \cite{bov01}; \cite{jan03}),
linear area technique (\cite{tar90}; \cite{law91};
\cite{sch92}; \cite{bal93}; \cite{meu99}),
monofractal box-counting technique (\cite{sta97};
\cite{gal98}; \cite{geo02}),
multifractal box-counting technique (\cite{law93};
\cite{cad94};\cite{law96}; \cite{con08})
multifractal wavelet technique: (\cite{hew08}).
This yields fractal dimensions for the 2d photospheric ARs ($D_{2d}$) that vary from $\sim1.1$ up to $\sim2.0$,
depending on the method used and the AR sample under consideration.

The fractal dimension is also useful because it provides a reliable test of various theoretical models and
simulations against observations. Cellular automata models (\cite{isl00,vla02}), percolation models (\cite{sei96,fra04}), random walk
diffusion models (\cite{law93}), and photospheric magnetoconvection simulations (\cite{jan03})
have been tested against observations through their fractal dimension.

The majority of the above-mentioned fractal analyses has been implemented in $2d$ and
applied to (mainly photospheric) magnetograms. Solar energetic events, however, take place above the photosphere.
Because of the turbulent photospheric motions, the magnetic helicity, and shear generated in the
coronal magnetic fields lead to magnetic
discontinuities, which can give rise to nanoflares, microflares, flares, and CMEs when some critical threshold is exceeded.

It would therefore be very interesting to seek the fractal dimension $D_{3d}$ describing the complexity
of the magnetic field in the $3d$ volume above a given $2d$ magnetogram. Aschwanden and Aschwanden (\cite{asc08a,asc08b}) were among the
first to follow this concept. They calculated the $D_{2d}$
fractal dimension of 20 flares, using a standard box-counting technique on data from the Transition Region and Coronal Explorer (TRACE).
To infer the $D_{3d}$ in the corona, they created an analytical flare geometry
model. The coronal arcade geometry includes three free parameters (arcade
length, width, and heliographic longitude), and  makes a number of
simplifying assumptions, such as (1) the arcade is near the equator
and latitudinal projection effects are neglected, (2) the magnetic
shear along the neutral line is neglected, (3) the neutral line is oriented
in the east-west direction, (4) individual loops in the arcade
are semicircular, and (5) the arcade length is assumed to be commensurable
with the arcade width when observed near the limb. This work has shown that $D_{2d}$ and $D_{3d}$
do not scale as $D_{2d}/2=D_{3d}/3$, as expected for isotropic structures,
but obey a more complex relation.

The purpose of the present work is to elaborate the relation between $D_{2d}$ and $D_{3d}$ further
by using - for the first time - magnetic field measurements and extrapolations.
We investigate whether a correlation between $D_{2d}$
and $D_{3d}$ actually exists and we have created an extended database of 38 ARs in the form of
vector magnetograms. Our input magnetograms come both from the Imaging Vector Magnetograph (IVM; Mickey et al., 1996; LaBonte at al., 1999)
of the  Mees Solar Observatory and
the spectropolarimeter (SP; \cite{lit01}) of the solar optical telescope (SOT) onboard Hinode.
We first calculated the $D_{2d}$ of the 2d magnetic structures
captured by the magnetograms through a conventional box-counting algorithm. To infer the $D_{3d}$ of
the volume above a given AR, we extrapolated the observed photospheric magnetograms into the corona. The extrapolation method-of-choice was the nonlinear force-free optimization method of Wiegelmann (2004), extending the work of Wheatland et al., (2000). Although nonlinear force-free extrapolation is  still a fully open research topic, Wiegelmann's (2004) method has been recognized by comparison studies involving several nonlinear force-free  extrapolation methods as one of the most reliable techniques (e.g. \cite{sch06}, \cite{met08}). Potential extrapolation was also used as a zero level of approximation for reasons of comparison. After the extrapolation, we identified unstable volumes (UnVos) in the corona as ensembles of adjacent magnetic discontinuity sites. The $3d$ fractal dimension $D_{3d}$ is determined as the box-counting fractal dimension of these UnVos.

This work is structured as follows. Section 2 describes the data used in this study along with the necessary corrections imposed
on them. Section 3 discusses the box counting and extrapolation techniques applied to our dataset, aiming to reveal correlations between 2d and 3d magnetic domains. Important parameters and quantities used are explained in detail,
whereas work-around solutions on standard methodology drawbacks are suggested. Section 4 presents our results and discusses
our findings. Finally, Sect. 5 summarizes our conclusions.


\section{Dataset}

   Nonlinear force-free extrapolation techniques typically require vector
   magnetograms that are not as widely available as conventional line-of-sight
   magnetograms. Here we have created a database of 38 different AR
   vector magnetograms from the IVM and Hinode's SOT/SP.

   The IVM obtains Stokes images in photospheric lines with $7 pm$ spectral resolution, $1arcsec$ spatial resolution
   ($\sim{0.55}$ arcsec per pixel in full resolution) over a field of $4.7 arcmin^{2}$ and polarimetric precision of $0.1\%$ (\cite{mic96}).
   We used both fully-inverted and quicklook IVM data. The quicklook data were obtained from the IVM Survey Data archive, made available at $http://www.cora.nwra.com/\sim{ivm}/IVM$-$SurveyData/$. The quicklook data reduction differs from the complete inversion in that it uses a
   simplified flat-fielding approach, takes no account of scattered or parasitic light, and no correction is
   attempted for seeing variations that occur during the data acquisition. On the other hand,
   the spectro polarimeter (SP) of SOT onboard Hinode obtains line profiles of two magnetically sensitive $Fe$ lines
   at $630.15$ and $630.25 nm$, using a $0.16\times 164$ $arcsec$ slit and has a spectral resolution of 30 mA (\cite{shi04}).

   In this study we used 10 fully inverted and 16 quicklook IVM vector magnetograms with 12 Level1D SOT magnetograms.
   To remove the intrinsic azimuthal ambiguity of $180^{o}$ we used the Non-Potential magnetic Field Calculation (NPFC)
   method of Georgoulis (\cite{geo05}). For computational convenience we also rebinned the disambiguated magnetograms into a 128x128 grid.

   That
   our input data derive from instruments with different spatial resolutions should not influence our analysis,
   as the fractal dimension is scale-invariant and does not depend on the spatial resolution of the instrument used.
   This is true for mathematical fractals however, because fractality holds for a finite range of sizes in realistic situations.
   For this reason, we first assume that all multi-instrument vector magnetograms belong to the same dataset, and then we duplicate the analysis using three different sets: one including fully-inverted IVM magnetograms, one including quicklook IVM data, and one including the Hinode/SOT data.


   \begin{figure}
   \centering
  \includegraphics[width=0.5\textwidth]{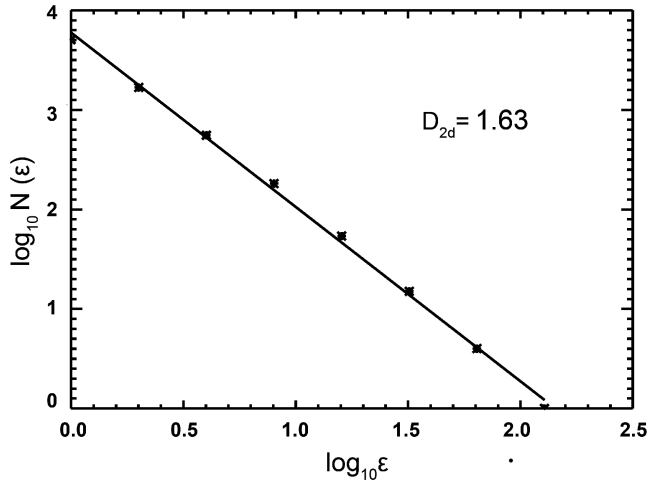}
   \caption{Logarithmic plot of $N(\epsilon)$ versus $\epsilon$ for AR 10953,
   where $D_{2d}=1.63$.}
              \label{FigVibStab}%
    \end{figure}


\section{Data analysis method}

Our analysis consists of 3 distinct steps. First we apply a box-counting algorithm
to the $128\times 128$ disambiguated photospheric magnetograms. We estimate the $D_{2d}$ fractal dimension
from the pixels with magnetic field strength exceeding a given threshold. Second, we apply
the Wiegelmann optimization algorithm to our vector magnetograms in order to nonlinearly extrapolate the magnetic field
from the photospheric boundary. We thus construct a 3d $128\times 128\times 128$ cube, within which the
magnetic field is unambiguously determined. We also perform potential extrapolation as a zero level of approximation.
Third, we calculate $D_{3d}$ via box-counting the sites within our cubic grid
which exceed a threshold of either the gradient of the magnetic field strength or the magnitude of the rotation (curl) of the magnetic field vector. The UnVos in the coronal volume are determined in this way. Finally, we investigate whether there is any correlation between
the inferred $D_{2d}$ and $D_{3d}$. The quality of the shown correlations is being judged by both the linear (Pearson) correlation coefficients.


\subsection{$D_{2d}$ Determination}

To calculate the $D_{2d}$ of each AR on the photospheric level,
we apply a standard box-counting technique, similar to McAteer et al. (2005).
As input data we use the disambiguated $128\times 128$ vector magnetograms (see Sect. 2).
The quantity examined is the magnetic field magnitude $|{\vec{B}}|$ at the photospheric level ($z=0$),
which can be unambiguously determined by the known magnetic field components $\vec{B_{x}}, \vec{B_{y}}, \vec{B_{z}}$
on the heliographic plane. A site is taken into account in the box-counting if its magnetic field magnitude exceeds
a specific threshold $B_{cr}$. If the number, $N$, of strongly magnetized
square boxes ($|{\vec{B}}|\geq B_{cr}$) scales with box size, $\epsilon$, as

   \begin{equation}
      N(\epsilon)\propto \epsilon^{-D_{2d}}
   \end{equation}then $D_{2d}$ is the fractal dimension, whereby $\epsilon$ is successively increased as $\epsilon=2^{n}, n = 0, 1, 2,...7 $,
 with maximum $\epsilon_{max}=128$. In practice, $D_{2d}$ is the scaling index of the power-law least-squares best fit between
 $N(\epsilon)$ and $\epsilon$. The criterion $|{\vec{B}}|\geq B_{cr}$ for identifying the photospheric ARs has been used widely in the literature (e.g. \cite{mca05}).

However, the above-mentioned methodology is known to suffer from two major problems. First and foremost, it is
sensitive to the selection of the critical threshold value $B_{cr}$. To determine a non-arbitrary
value of $B_{cr}$, we apply a histogram method, by constructing the histogram of the
field magnitudes of all magnetograms in our database.
We then fit a Gaussian to this histogram and define $B_{cr}$
as the value, above which the histogram deviates from the Gaussian.
This test yields $B_{cr}=230\,G$.
The second drawback lies in the least-squares best fit being prone to large errors,
mainly due to the small number of data-points in its dynamical range. To overcome this problem,
we apply a goodness of fit test in the form of a chi-square test, following Isliker (\cite{isl92}).
The chi-square test is applied to a sliding window on the linear representation of the
scaling relation, and it indicates with a 90\% level of significance whether a range
of the overall scaling is indeed a power law.
In addition,
we demand that the power-law
scaling extends over at least one order of magnitude,
in order to
yield sufficient dynamical range for a reliable estimate
of the fractal dimension. As an example in Fig.1
we show the plot of $\log N_{\epsilon}$ versus $\log \epsilon$ for AR 10953,
where $D_{2d}=1.63$.

   \begin{figure*}
   \centering
  \includegraphics[width=1.0\textwidth]{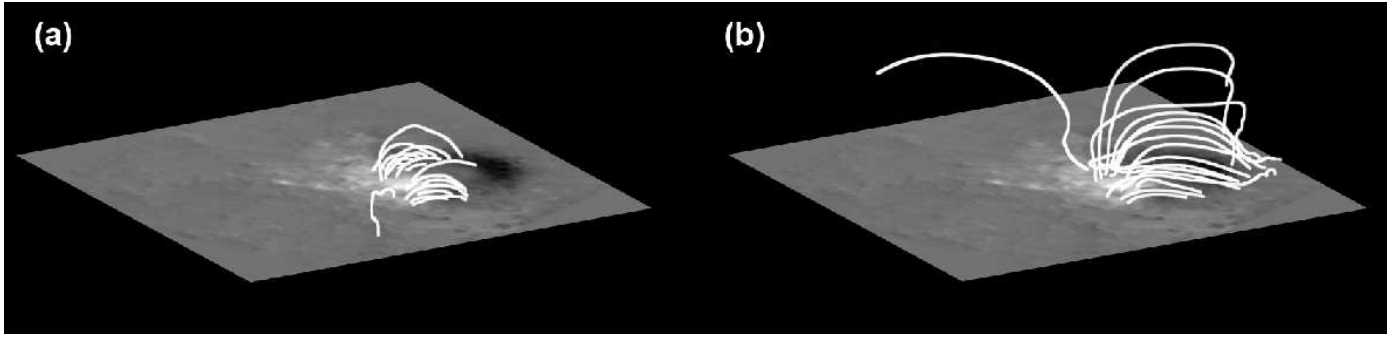}
   \caption{a) Potential extrapolation for AR 10953. b) The respective NLFF
                field solution, where the boundary vector magnetogram has been preprocessed.}
              \label{FigGam}%
    \end{figure*}


\subsection{Nonlinear force-free extrapolation}

The next step is to extrapolate the photospheric magnetic fields.
Potential extrapolation is the lower limit (zero level of) approximation and it provides the simplest
force-free magnetic configuration. As such, it is expected to yield the best correlation
between $D_{2d}$ and $D_{3d}$. A linear, but non-potential, force-free field extrapolation
is the first level of approximation, while the nonlinear force-free (NLFF) field extrapolation
is the second level of approximation. The most realistic treatment would be a non-force-free static
extrapolation or a magnetohydrostatic / magnetohydrodynamic model using the photospheric fields
as boundary conditions. The latter, however, implies analysis and computational resources that
far exceed the scope of this work. We chose the NLFF field extrapolation for two reasons:
\begin{enumerate}
	 \item While the linear force-free assumption may have some validity in a minimum-energy AR corona,
           it cannot be trusted in the chromosphere or, even more, in the AR photosphere. There the
           relatively high value of the plasma beta-parameter implies Lorentz forces that cannot be neglected.
  \item  Because the purpose of this work is to investigate whether a correlation between $D_{2d}$ and $D_{3d}$ exists,
         it is crucial to refrain from using an extrapolation that is susceptible to artificial correlations between
         the 2d and the 3d domains. A linear extrapolation method would bind the 3d coronal magnetic field to its
         photospheric 2d boundary.
\end{enumerate}

The NLFF extrapolation used here is based on the optimization technique introduced by
Wheatland et al. (2000) and further developed by Wiegelmann (\cite{wie04}; \cite{wie06}; \cite{wie08}).
This technique reconstructs force-free magnetic
fields from their boundary values, based on minimizing the Lorentz force and the divergence of the magnetic field vector in the extrapolation volume:

   \begin{equation}
      L=\int_{V}w(x,y,z)[|\vec{B}|^{-2}|(\vec{\nabla}\times\vec{B})\times\vec{B}|^{2}+|\nabla\cdot\vec{B}|^{2}]d^{3}x
   \end{equation}.

In the above functional,  $w(x,y,z)$ is a weighting function and $V$ denotes the extrapolation volume.
A force-free state is reached when $L=0, w>0$. For $w(x,y,z)=1$,
the magnetic field must be available on all 6 boundaries of our cubic box for the optimization algorithm to work.
Nevertheless, real vector magnetograms provide the magnetic field only for the bottom boundary,
whereas the edge top and lateral magnetic field values remain unknown. The weighting function is thus used to
reduce the dependence of the interior solution on the unknown boundaries. We introduced a buffer boundary
of $100$ grid points towards the lateral and top boundaries of the computational box. We then chose $w(x,y,z)=1$ in
the inner, physical domain and let $w$ drop to 0 with a cosine-profile in the buffer boundary towards the lateral and top
boundaries of the computational box (see \cite{wie04} for details).

An additional useful attribute of Wiegelmann's NLFF field extrapolation code is the preprocessing alternative it offers. As the
photospheric magnetic field is in principle inconsistent with the force-free approximation, a preprocessing procedure was
developed by Wiegelmann et al. (2006) in order to drive NLFF fields closer to a force-free equilibrium. Preprocessing minimizes the forces and torques in the system thus satisfying the force-free requirements more closely. While preprocessing the photospheric magnetograms, however, we applied some smoothing to the field components. To study the effects of this smoothing we duplicated the fractal-dimension calculation for unpreprocessed photospheric fields, as well. As shown in Sect. 4, preprocessing - including smoothing - does not affect $D_{2d}$ significantly.

Figure 2 depicts the field lines of the extrapolated magnetic field for AR 10953, as calculated by the optimization algorithm.
Frame a) shows the current-free (potential) extrapolation for reference while the respective NLFF field solution is shown in Frame b).


\subsection{$D_{3d}$ Determination}

The fractal dimension $D_{3d}$ in the extrapolation volume is determined by means of the box-counting and
goodness-of-fit methods described in Section 3.1. What differs in this
case is the quantity used to determine which grid sites in the volume will be taken into account. This quantity will determine
the UnVos present in the cubic grid. We have two different ways of calculating these UnVos:
\begin{enumerate}
  \item The average magnetic field gradient $G_{av}$\\
           For every site $i,j,k$ within our grid, we calculate the average magnetic field gradient with its neighboring sites as\\

           $G_{av_{i,j,k}}=\frac{|\vec{B_{i,j,k}}-\vec{B_{sum}}/nn|}{|B_{i,j,k}|}$\\
           $\vec{B_{sum}}=(B_{sumx},B_{sumy},B_{sumz})$.\\
           We define:\\
           $B_{sumx}=B_{x_{i+1,j,k}}+B_{x_{i-1,j,k}}+B_{x_{i,j+1,k}}+B_{x_{i,j-1,k}}+B_{x_{i,j,k+1}}+B_{x_{i,j,k-1}}$\\
           $B_{sumy}=B_{y_{i+1,j,k}}+B_{y_{i-1,j,k}}+B_{y_{i,j+1,k}}+B_{y_{i,j-1,k}}+B_{y_{i,j,k+1}}+B_{y_{i,j,k-1}}$\\
           $B_{sumz}=B_{z_{i+1,j,k}}+B_{z_{i-1,j,k}}+B_{z_{i,j+1,k}}+B_{z_{i,j-1,k}}+B_{z_{i,j,k+1}}+B_{z_{i,j,k-1}}$.\\

           Depending on the location of each site within the volume, the number of nearest neighbors $nn$ assumes the values
           $nn=3,4,5,6$. The physical explanation
           for selecting this criterion lies in a steep gradient of the magnetic field strength being thought
           to favor magnetic reconnection in $3d$, in the absence
           of null points (\cite{pri03}).

\item The normalized magnetic field curl $C_{n}$\\
           For every site $i,j,k$ within our grid we calculate the normalized magnetic field curl as\\
           $C_{n_{i,j,k}}=|\frac{\vec{\nabla\times \vec{B_{i,j,k}}}}{B_{i,j,k}}|$.\\

           This criterion will obviously highlight areas of high electric-current concentrations, which are known
           to play a role in
           the formation of magnetic instabilities.
\end{enumerate}

During the application of the box-counting method, we consider a site as unstable if it exceeds a critical threshold value
(gradient or curl respectively). This critical value is determined with the histogram method described in
Sect. 3.1, yielding $G_{cr}=0.09$ for the average gradient criterion and $C_{cr}=0.005$ for the normalized curl one.
If the number, $N$, of unstable square boxes ($G_{av}\geq G_{cr}$ or $C_{n}\geq C_{cr}$) scales with box size, $\epsilon$, as

   \begin{equation}
      N(\epsilon)\propto \epsilon^{-D_{3d}}
   \end{equation} then $D_{3d}$ is the 3d fractal dimension, whereby $\epsilon$ is once again successively
increased as $\epsilon=2^{n}, n = 0, 1, 2,...7 $, with maximum
 $\epsilon_{max}=128$. As an example in Fig.3
we show the logarithmic plot of $N(\epsilon)$ versus $\epsilon$ for AR 10953,
where $D_{3d}=2.52$ for the gradient (frame a) and $D_{3d}=2.91$ for the curl (frame b).
In both cases preprocessing were conducted prior to the NLFF extrapolation.

   \begin{figure*}
   \centering
   \includegraphics[width=1.0\textwidth]{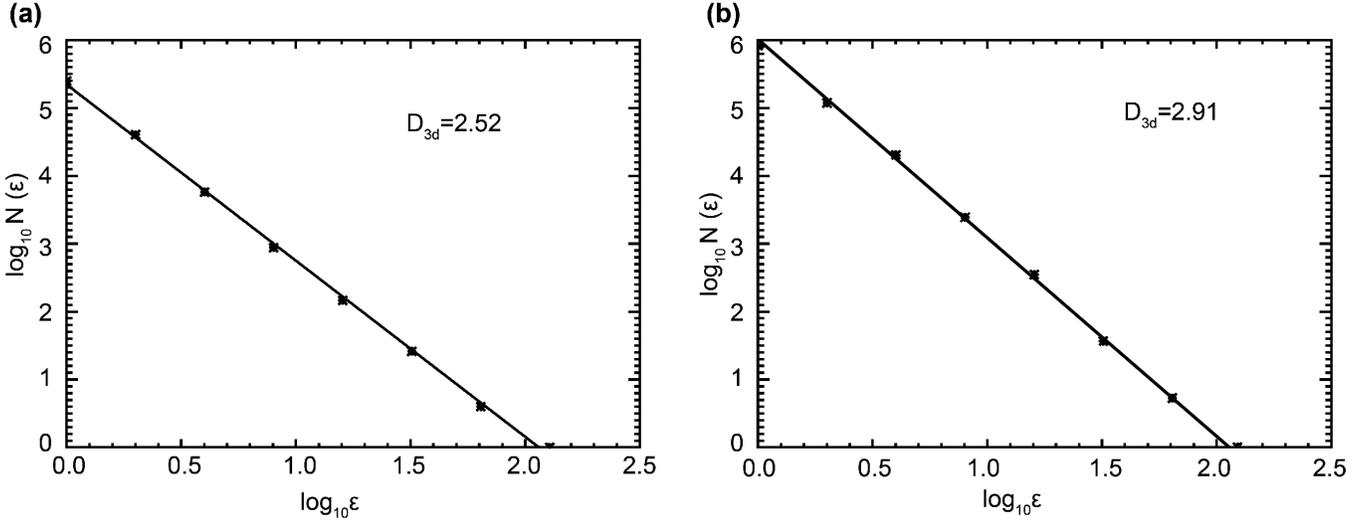}
   \caption{Logarithmic plot of $N(\epsilon)$ versus $\epsilon$ for the UnVos in AR 10953.
             UnVos have been calculated using $G_{av}$ (Frame a) and $C_{n}$ (Frame b).
             The respective fractal dimensions $D_{3d}$ are also shown.}
             \label{FigGam}%
    \end{figure*}


\section{Results}

As discussed in Sect. 3.2, several combinations were investigated with respect to the extrapolation methods and correlation coefficients.
We first present the case of NLFF extrapolation with preprocessing. This is carried out
before NLFF extrapolation, as unpreprocessed magnetograms are not
consistent with force-free extrapolations.
Consequently it would be very hard to approximate a nonlinear force-free equilibrium
consistent with unpreprocessed data.

First we investigate the relation between $D_{2d}$ and $D_{3d}$, when the latter is calculated through
the average ($G_{av}$) magnetic field gradient. The corresponding Pearson correlation coefficient between $D_{2d}$ and $D_{3d}$ is $-0.154$.
The corresponding probabilities are shown in Table 1 for Pearson coefficient.
We then use the normalized magnetic field curl ($C_{n}$),
to calculate $D_{3d}$. In this second attempt the Pearson correlation coefficient between $D_{2d}$ and $D_{3d}$
is $0.251$, as shown along with the corresponding probabilities in Table 1.
Evidently, when NLFF extrapolation is used, there is no correlation between $D_{2d}$ and $D_{3d}$ at the $95\%$ significance level
and this result is independent of
the criterion used to quantify the UnVos. Moreover, independently of the magnetic complexity of an AR in the photosphere
(as quantified by $D_{2d}$), the complexity of the generated UnVos varies within a more or less well-defined range, as can be seen
in Table 2. The absence of correlation
between $D_{2d}$ and $D_{3d}$, as well as the range of $D_{3d}$ values, is shown in frame a) of Fig.4, where the NLFF case is presented.
This figure suggests that - although the complexity in the photosphere may vary significantly - the UnVos in the corona retain a more or less
well-defined behavior.

If we attempt to locate these
UnVos within the $128\times 128 \times 128$ volume, one finds that more than $80\%$ of them are
accumulated in the lower layers, whereas only $20\%$ are found at heights $15\leq z\leq 30$, in units of the
boundary magnetogram's linear pixel size.
The magnetic discontinuities are restricted to the lower coronal layers,
whereas only a few weak additional discontinuities are identified in higher $z\geq30$ layers.
This result is even better illustrated if we
slice our volume in layers along the $z$ axis and mark which sites per layer host magnetic instabilities. Figure 5
shows the results for AR 10953, where $G_{av}$ is used to determine UnVos in heights $z=17,22,27,32$
correspondingly. It is obvious that the higher they are from the photosphere, the less and the weaker they are.
This can be shown alternatively by means of the fractal dimension: we identify
which sites per layer $z_{o}$ satisfy the relation $G_{av_{i,j,z_{o}}}\geq G_{cr}$
and by using the box counting technique we calculate the fractal dimension $D^{'}_{2d_{z_{o}}}$.
The distribution of $D^{'}_{2d}$ with height is shown in Fig. 6. It is evident that each layer for
$0\leq z\leq 20$ is almost completely filled by UnVos, yielding a nearly Euclidian (non-fractal) dimension of $D^{'}_{2d}\simeq 2$.
For heights $20\leq z\leq 30$, $D^{'}_{2d}$ decreases from 2 to 1, while for $z>30$ $D^{'}_{2d}$ becomes smaller than 1,
indicating very small and isolated, "dust-like", UnVos.
   \begin{figure*}
   \centering
   \includegraphics[width=1.0\textwidth]{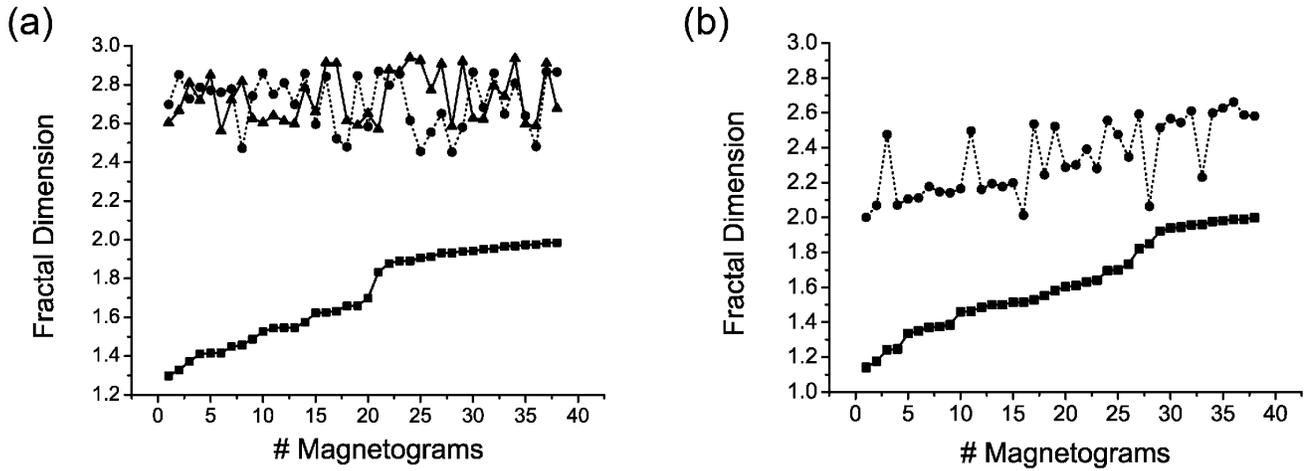}
   \caption{Fractal dimensions $D_{2d}$ and $D_{3d}$ for all 38 magnetograms in our sample. Frame a)
   depicts the case of NLFF extrapolation with preprocessing, whereas frame b) depicts the case of potential extrapolation.
   In both Frames a) and b) the solid line with squares
   represents $D_{2d}$ in ascending order and the dotted line with circles stands for $D_{3d}$ when UnVos are identified by the $G_{av}$ criterion,
   In Frame a) the solid line with triangles reflects $D_{3d}$ when UnVos are identified based on the $C_{n}$ criterion. }
              \label{FigGam}%
    \end{figure*}


This finding corroborates previous results. Regnier \& Priest (\cite{reg07}) show that there is a clear preference for
free energy accumulation very close to the photospheric level. Given this physical property, we attempted an
alternative approach by omitting the layers that are filled by UnVos and
focusing only on the part of the volume where the filling with UnVos starts becoming sparse.  In this case, would there be any
correlation between the photospheric driver and the higher coronal structures? To examine this possibility,
we calculated $D_{3d}$ starting from the specific $z$, above which only the remaining $20\%$ of UnVos is identified, thus excluding
the lower coronal layers with a large filling factor.
This leads to a decrease in the fractal dimension $D_{3d}$, as expected, and an increase in the correlation coefficients,
but we still find a lack of any significant correlation between $D_{2d}$ and $D_{3d}$.

   \begin{figure*}
   \sidecaption
  \includegraphics[width=12cm]{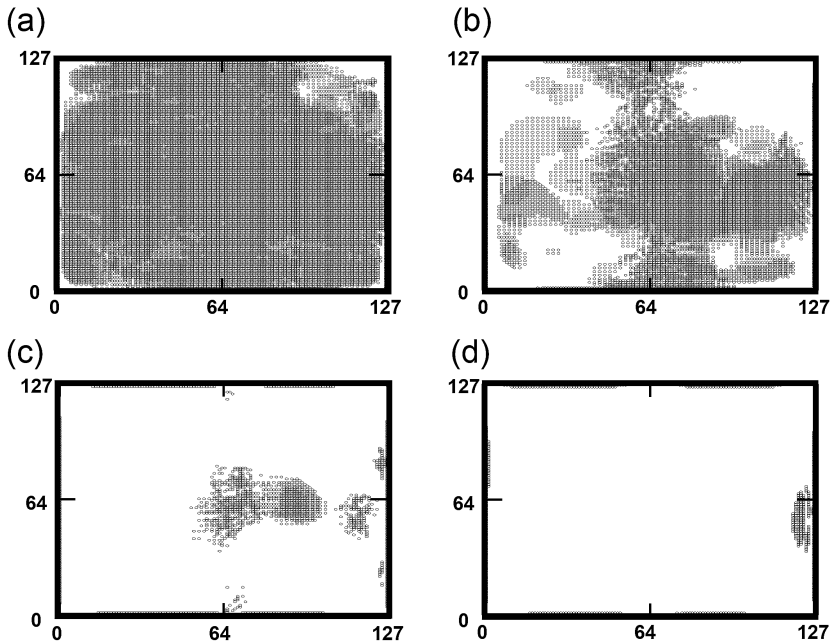}
   \caption{Sites per $z$ layer that host magnetic instabilities according to the $G_{av}$ criterion for AR 10953. a) $z=17$, b) $z=22$, c) $z=27$, d) $z=32$.}
              \label{FigGam}%
    \end{figure*}

   \begin{figure}
   \centering
  \includegraphics[width=0.5\textwidth]{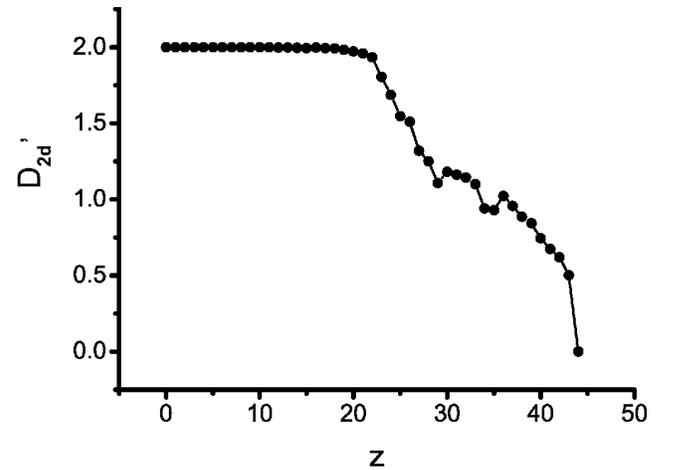}
   \caption{Fractal dimension $D^{'}_{2d}$ versus height $z$ for AR 10953.}
              \label{FigGam}%
    \end{figure}

%
\begin{table}
\caption{Pearson correlation coefficients between $D_{2d}$ and $D_{3d}$ for the potential and the NLFF cases.
}             
\label{table:1}      
\centering                          
\begin{tabular}{c c c c}        
\hline                
Extrapolation  & Used 3d   & Correlation & Corresponding   \\    
Method & criterion  & with $D_{2d}$& Probability   \\    
\hline                        
   Potential & $G_{av}$ & 0.719 & 99\%\\
   NLFF with preprocessing & $G_{av}$ & -0.154 & 65\%\\
   NLFF with preprocessing & $C_{n}$ & 0.251 & 87\%\\
\hline                                   
\end{tabular}
\end{table}
%

%

%

%
\begin{table}
\caption{Range of $D_{2d}$ and $D_{3d}$ values, depending on the criterion used.
$100\%$ of UnVos are considered.  }            
\label{table:3}      
\centering                          
\begin{tabular}{c c c c}        
\hline                
Extrapolation & Fractal & Used & Range of   \\    
Method & Dimension & criterion  & values  \\    
\hline                        
Potential  & $D_{2d}$ & $B_{cr}$    & $1.57\pm 0.43$ \\
Potential   &   $D_{3d}$ & $G_{av}$     & $2.331\pm 0.331$ \\
NLFF with preprocessing &  $D_{2d}$ & $B_{cr}$     & $1.64\pm 0.34$ \\
NLFF with preprocessing   &   $D_{3d}$ & $G_{av}$     & $2.66\pm 0.21$ \\
NLFF with preprocessing   &   $D_{3d}$ & $C_{n}$    & $2.75\pm 0.19$ \\
\hline                                   
\end{tabular}
\end{table}
%

%
%

As the first row of Table 1 shows, the only case where we see a considerable correlation at the level of $95\%$ significance between $D_{2d}$ and $D_{3d}$ is when we use potential field extrapolation. Figure 4b illustrates the correlation between $D_{2d}$ and $D_{3d}$ when the magnetic gradient criterion is used in the potential case. This is a reasonable result, considering that the potential extrapolation produces the simplest force-free
magnetic configuration. At this zero level of approximation the significant correlation revealed between the photospheric and coronal structures is attributed to the lack of currents. Similar results can also be reproduced for the linear force-free extrapolation: the correlation between $D_{2d}$ and $D_{3d}$ is again significant at the level of $99\%$, with a Pearson correlation coefficient of 0.75. Inspecting Fig. 4 and Table 2, we also notice that $D_{3d}$ is larger for NLFF field extrapolations compared to potential field extrapolations. This means more extended UnVos, hence a higher spatial filling in the NLFF case, as expected. Other than that, UnVos show a strong preference for accumulating at low altitudes in the extrapolation volume for both NLFF and potential fields. Similar results were found by Vlahos \& Georgoulis (2004),
where linear force-free extrapolation is used. Their study does not apply a fractal analysis but identifies many magnetic discontinuities whose free magnetic energies and volumes obey well-formed power-law distribution functions.

Furthermore, we investigated whether the smoothing imposed by the preprocessing has severely affected the magnetic field. For this purpose
we compared the fractal dimensions $D_{2d}$ of the observed (unpreprocessed) and the preprocessed magnetograms in the photospheric boundary.
As shown in Table 3, the fractal dimension in the photospheric level is not significantly altered by preprocessing. The correlation between the preprocessed and raw magnetic data is significant at a level of $99\%$, as also illustrated in Fig. 7. While no evidence of correlation between $D_{2d}$ and $D_{3d}$ exists in the NLFF field limit, significant correlation exists in the potential limit, even without preprocessing.

%
\begin{table}
\caption{Pearson correlation coefficient of the fractal dimensions $D_{2d}$ between the unpreprocessed
and the preprocessed photospheric magnetograms.}             
\label{table:4}      
\centering                          
\begin{tabular}{c c c}        
\hline                
    & Correlation   & Corresponding   \\    
    & Coefficient   & Probability   \\    
\hline                        
Pearson & 0.863 & $99\%$ \\
\hline                                  
\end{tabular}

\end{table}

   \begin{figure}
   \centering
  \includegraphics[width=0.5\textwidth]{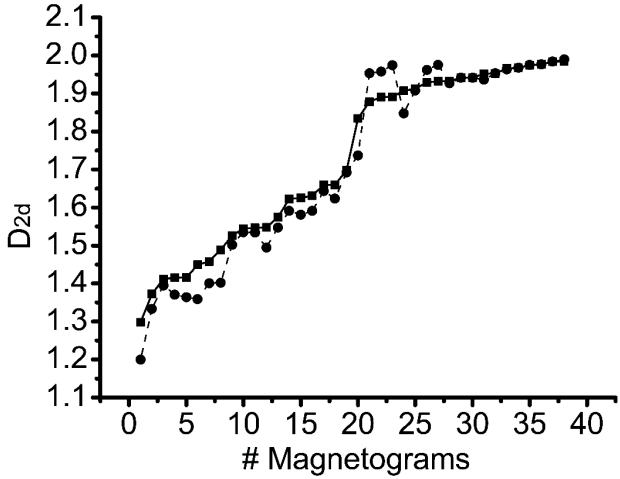}
   \caption{Fractal dimensions $D_{2d}$ for raw (dashed line with circles) and preprocessed (straight line with squares) magnetograms.}
              \label{FigGam}%
    \end{figure}

Finally, we investigated how the spatial resolution differences between the datasets derived from various instruments influence
our results. We separated the calculated fractal dimensions ($D_{2d}$ and $D_{3d}$) into three groups, depending on the instrument they
are derived from. This separation substantially decreases the number of data in each sample and is prompt to influence our statistical
results. Nevertheless, this is the only way to investigate how different spatial resolution and inversion sophistication (in case of IVM data) can affect the analysis. Poor statistics will lead to relatively low confidence levels. To properly define the threshold values for
each of the subsets, we applied the histogram test described in Sects. 3.1 and 3.2 for 2d and 3d correspondingly separately to each subset; nevertheless, the derived threshold values in all cases were very close to the global $B_{cr}$, $G_{cr}$, and $C_{cr}$ values found for all 38 magnetograms, thus allowing us to retain them also for all separate subsets.

Table 4 corresponds to Table 1 but only contains the fully inverted data derived from IVM (10 magnetograms).
Table 5 corresponds to Table 1 but only contains the data derived from HINODE (12 magnetograms).
Table 6 corresponds to Table 1 but only contains the quicklook data derived by IVM (16 magnetograms).
Finally, Table 7 corresponds to Table 3 but for each distinct dataset separately.
 Although substantial differences exist between the various datasets, it is evident that the results found for the total dataset
 are qualitatively retained also for the subsets of data coming from
 different instruments. The correlation between $D_{2d}$ and $D_{3d}$ is significant for potential extrapolation, whereas it is absent in
 the NLFF limit. Preprocessing does not significantly alter the photospheric magnetic fields in any case.

%
\begin{table}
\caption{Fully Inverted IVM subset: Pearson correlation coefficients between $D_{2d}$ and $D_{3d}$ for the potential and the NLFF cases.
}             
\label{table:5}      
\centering                          
\begin{tabular}{c c c c}        
\hline                
Extrapolation  & Used 3d   & Correlation & Corresponding   \\    
Method & criterion  & with $D_{2d}$& Probability   \\    
\hline                        
   Potential & $G_{av}$ & 0.758 & 99\%\\
   NLFF with preprocessing & $G_{av}$ & -0.278 & 61\%\\
   NLFF with preprocessing & $C_{n}$ & -0.102 & 25\%\\
\hline                                   
\end{tabular}
\end{table}
\begin{table}
\caption{HINODE subset: Pearson correlation coefficients between $D_{2d}$ and $D_{3d}$ for the potential and the NLFF cases.
}             
\label{table:8}      
\centering                          
\begin{tabular}{c c c c}        
\hline                
Extrapolation  & Used 3d   & Correlation & Corresponding   \\    
Method & criterion  & with $D_{2d}$& Probability   \\    
\hline                        
   Potential & $G_{av}$ & 0.919 & 99\%\\
   NLFF with preprocessing & $G_{av}$ & -0.5 & 91\%\\
   NLFF with preprocessing & $C_{n}$ & 0.427 & 85\%\\
\hline                                   
\end{tabular}
\end{table}
\begin{table}
\caption{Quicklook IVM subset: Pearson correlation coefficients between $D_{2d}$ and $D_{3d}$ for the potential and the NLFF cases.
}             
\label{table:8}      
\centering                          
\begin{tabular}{c c c c}        
\hline                
Extrapolation  & Used 3d   & Correlation & Corresponding   \\    
Method & criterion  & with $D_{2d}$& Probability   \\    
\hline                        
   Potential & $G_{av}$ & 0.765 & 93\%\\
   NLFF with preprocessing & $G_{av}$ & 0.238 & 65\%\\
   NLFF with preprocessing & $C_{n}$ & 0.099 & 30\%\\
\hline                                   
\end{tabular}
\end{table}
\begin{table}
\caption{Pearson correlation coefficients of the fractal dimensions $D_{2d}$ between the unpreprocessed
and the preprocessed photospheric magnetograms in the distinct data subsets.}             
\label{table:7}      
\centering                          
\begin{tabular}{c c c c}        
\hline                
    & Data  & Correlation   & Corresponding   \\    
    & Subset & Coefficient   & Probability   \\    
\hline                        
Fully Inverted IVM & Pearson & 0.770 & $99\%$ \\
HINODE & Pearson & 0.989 & $99\%$ \\
Quicklook IVM & Pearson & 0.688 & $95\%$ \\
\hline                                  
\end{tabular}

\end{table}

\section{Discussion and conclusions}

This study investigates whether the complexity of the photospheric
magnetic field correlates with the complexity of the UnVos formed in the corona.
Using 38 vector magnetograms, we

\begin{itemize}
	 \item estimate the fractal dimension $D_{2d}$
from the pixels with a magnetic field strength exceeding a given threshold $B_{cr}$ at the photospheric level through a standard box-counting method
  \item extrapolate the magnetic field from the photospheric boundary using a
   \begin{enumerate}
   	 \item nonlinear force-free optimization algorithm with preprocessing at the photospheric level,
     \item standard potential extrapolation algorithm
   \end{enumerate}
  \item calculate $D_{3d}$ by box-counting the sites within our cubic grid, which exceeds a threshold in
\begin{enumerate}
	 \item the averaged magnetic field gradient $G_{av}$ with respect to their neighboring sites,
     \item the normalized magnetic field curl $C_{n}$ (only for the NLFF fields)
\end{enumerate}
  \item investigate whether there is any correlation between
the inferred $D_{2d}$ and $D_{3d}$.
\end{itemize}

Our results show that there is no correlation revealed between $D_{2d}$ and $D_{3d}$ for the NLFF case.
The spatial distribution of the UnVos with height shows that $\geq 80\%$ of the magnetic discontinuities are accumulated in the
lower corona (within $20$ $Mm$ from the photosphere). The system is evidently highly unstable at these heights, yielding processes that are clearly
nonlinear. It is this strong nonlinearity at lower layers that does not allow the corona to respond proportionally to changes imposed by the photospheric driver or, conversely, does not allow the line-tied photosphere to strongly respond to a restructuring of the coronal magnetic fields caused by an energetic event (flare, CME). The results of Sudol \& Harvey (\cite{sud05})
establish that, even for the largest flares, the photospheric response is significant, but rather small (median value of photospheric magnetic field
change at 90 G) even given that these flares have to occur close to the photosphere where most of the free magnetic
energy resides.

In contrast to the results for NLFF fields, we have obtained a significant correlation between $D_{2d}$ and $D_{3d}$ in the case of potential field extrapolations. As already explained, this should be attributed to the absence of currents, which forces the coronal magnetic structures to closely follow the photospheric driver. Therefore, the currents in the NLFF extrapolated fields lead to a more complex magnetic topology and the loss of correlations with the photospheric fields.

It is interesting to notice the strong accumulation of UnVos close to the lower (photospheric) boundary in both the potential and the NLFF fields. We believe this is not an artifact and that it has to do with the fine, fibril structure of the photospheric magnetic fields, which gradually fades as we move toward the corona. The structure of these forced fields is long known (Livingston \& Harvey 1969; Howard \& Stenflo 1972; Stenflo 1973) and it becomes less prominent at higher layers until the magnetic field fills the entire coronal volume, excluding small isolated areas of current sheets and tangential discontinuities in general (Parker 2004 and references therein). UnVos are designed to indicate these discontinuities and, as such, they will strongly accumulate at lower layers. This will happen regardless of the extrapolation method that will tend to create smooth fields in the volume, at the same time accommodating the finely structured lower boundary.

Further indications regarding of the absence of correlation between the photospheric and the coronal processes and structures are provided by
Metcalf et al. (\cite{met94}), who examined the spatial and temporal relationship between coronal structures observed with
the Soft X-ray Telescope (SXR)
on board Yohkoh spacecraft and the vertical electric current density derived from photospheric vector magnetograms. Metcalf et al.
(\cite{met94}) found no evidence directly linking the electric currents observed in the photosphere to the heating of the coronal plasma indicated by the SXR brightness
and temperature.

The work of Aschwanden and Aschwanden (\cite{asc08a,asc08b}) - investigating the relationship between the fractal dimension
of flare images captured by TRACE (in $2d$) to the fractal dimension of coronal arcades produced by an analytical geometric
model (in $3d$) - indicates a complex relation between $D_{2d}$ and $D_{3d}$. From our results, even this relation may be destroyed when fewer simplifications are used and the forced photospheric fields are taken into account. That the photospheric fields include significant Lorentz forces has been shown by Metcalf et al. (1995) and Georgoulis \& LaBonte (2004).

Our own and previous independent results seem to support the conjecture that the absence of
correlation between the photospheric and coronal fractal dimensions would still be the case (in fact, correlation should probably become worse) if a more realistic static or dynamic non-force-free modeling of the coronal field was used. Photospheric
turbulence remains the driver for the coronal instabilities, but the strong nonlinearity
of the system in the lower coronal layers destroys any kind of direct relation between the
photospheric structures and their coronal counterparts. The photospheric
driver forces the system to accumulate a large number of magnetic discontinuities
that store enough energy to explain the statistical properties
of the solar activity in case of release. These discontinuities form
patterns that do not follow the morphological properties of the
photospheric magnetic flux concentrations, but have a
strong impact on the expected dynamical activity of the system,
namely, the magnetic energy release and the subsequent
particle acceleration processes (\cite{vla04b}).

\begin{acknowledgements}
      We are grateful to Thomas Wiegelmann for kindly contributing his NLFF extrapolation code to this analysis and to Bruce Lites who provided useful information regarding the Level 1D SOT/SP magnetograms from Hinode. Hinode is a Japanese mission developed and launched by ISAS/JAXA, with NAOJ as domestic partner and NASA and STFC (UK) as international partners. It is operated by these agencies in co-operation with ESA and NSC (Norway). IVM magnetograms were obtained by the staff of the U. of Hawaii Mees Solar Observatory, Air Force Office of Scientific Research contract F49620-03-C-0019. IVM quicklook data were obtained from work supported by the National Science Foundation under Grant No. 0454610. Any opinions, findings, and conclusions or recommendations expressed in this material are those of the authors and do not necessarily reflect the views of the National Science Foundation (NSF). Eva Ntormousi and Tassos Fragos are acknowledged
      for their constructive comments in the course of this work. Finally, we would like to thank the referees for their very constructive comments.
\end{acknowledgements}

\end{document}